\documentclass[journal]{IEEEtran}

\usepackage{cite,graphicx,amsmath,amssymb}
\usepackage{subfigure}
\usepackage{fancyhdr}
\usepackage{mdwmath}
\usepackage{mdwtab}
\usepackage{balance}
\usepackage{xcolor}
\usepackage{bm}

\usepackage{algorithm}
\usepackage{algorithmic}
\usepackage{multirow}
\usepackage{flafter}
\usepackage{comment}

\providecommand{\url} [1]{#1}

\newtheorem{theorem}{Theorem}

\begin{document}

\onecolumn

\title{Simultaneously Transmitting and Reflecting (STAR) Intelligent Omni-Surfaces, Their Modeling and Implementation}
%


\author{Jiaqi\ Xu, Yuanwei\ Liu, Xidong Mu, Joey Tianyi Zhou, Lingyang Song, H. Vincent Poor, and Lajos Hanzo.

\thanks{J. Xu and Y. Liu are with the School of Electronic Engineering and Computer Science, Queen Mary University of London, London E1 4NS, UK. (email:\{jiaqi.xu, yuanwei.liu\}@qmul.ac.uk).}
\thanks{X. Mu is with the School of Artificial Intelligence, Beijing University of Posts and Telecommunications, Beijing, 100876, China
(email: muxidong@bupt.edu.cn).}
\thanks{J. T. Zhou is with Institute of High Performance Computing, A*STAR, Singapore (email: zhouty@ihpc.a-star.edu.sg)}
\thanks{L. Song is with Department of Electronics, Peking University, Beijing, China (e-mail: lingyang.song@pku.edu.cn).}
\thanks{H. V. Poor is with the Department of Electrical Engineering, Princeton University, Princeton, NJ 08544 USA (e-mail: poor@princeton.edu).}
\thanks{L. Hanzo is with the school of Electronics and Computer Science, University of Southampton, Southampton SO17 1BJ, U.K.
(e-mail: lh@ecs.soton.ac.uk).}

}
\maketitle
\begin{abstract}
With the rapid development of advanced electromagnetic manipulation technologies, researchers and engineers
are starting to study smart surfaces that can achieve enhanced
coverages, high reconfigurability, and are easy to deploy. Among
these efforts, simultaneously transmitting and reflecting intelligent omni-surface (STAR-IOS) is one of the most promising categories. Although pioneering works have demonstrated the benefits of STAR-IOSs in terms of its wireless communication performance gain, several important issues remain unclear including
practical hardware implementations and physics-compliant models for STAR-IOSs. As a consequence, the feasibility of employing
STAR-IOSs in wireless networks is seriously put in doubt. In
this paper, we answer these pressing questions of STAR-IOSs by
discussing four practical hardware implementations of STARIOSs, as well as three hardware modelling methods and five
channel modelling methods. These discussions not only categorize
existing smart surface technologies but also serve as a physicscompliant pipeline for further investigating the STAR-IOSs.

\end{abstract}

\section{Introduction}

The increasing research interest in the topic of reconfigurable intelligent surfaces (RISs) and the stringent requirements of beyond-5G wireless communication networks underline the need for developing smart surfaces that are more flexible and powerful \cite{pan}.
The RIS, with its ability to control the phase shifts of its reflected signal purely relying on passive tunable parts, has been envisioned as a key enabler for the fledgling smart radio environment (SRE) concept~\cite{ahead}.
However, conventional reflecting-only RISs impose topological constraints on their deployment because they can only reflect signals to one side of the surface, while leaving any users located at its back side in service outage \cite{liu_survey}.

To address this limitation, there are an increasing number of research contributions that consider an alternative type of smart surface that allows wireless signals incident on either side of the surface to be simultaneously reflected and transmitted.
For instance, the authors of \cite{STAR_mag} proposed the simultaneously transmitting and reflecting RIS (STAR-RIS) concept for achieving 360 degrees SRE coverage.
In~\cite{IOS}, the authors introduced the concept of intelligent omni-surfaces (IOSs), which are capable of serving mobile users on both sides of the surface.

Various research contributions have demonstrated the superiority of simultaneously transmitting and reflecting intelligent omni-surfaces (STAR-IOSs\footnote{Since \textit{STAR-RISs} and \textit{IOSs} share similar concepts in the existing literature, in this work, we use the term STAR-IOS for consistency.}) over conventional reflecting-only RISs. For example, the authors of \cite{xu_star} demonstrated the extended coverage of STAR-IOSs by studying both the channel gains and diversity orders of users on both sides. In \cite{Xidong}, a convenient signal model and practical operating protocols were proposed for STAR-IOSs. The authors of \cite{shuhang} jointly optimized the STAR-IOS phase shift design and beamforming at the base station for maximizing the sum-rate of multiple users. 

\begin{figure*}[t!]
    \begin{center}
        \includegraphics[scale=0.4]{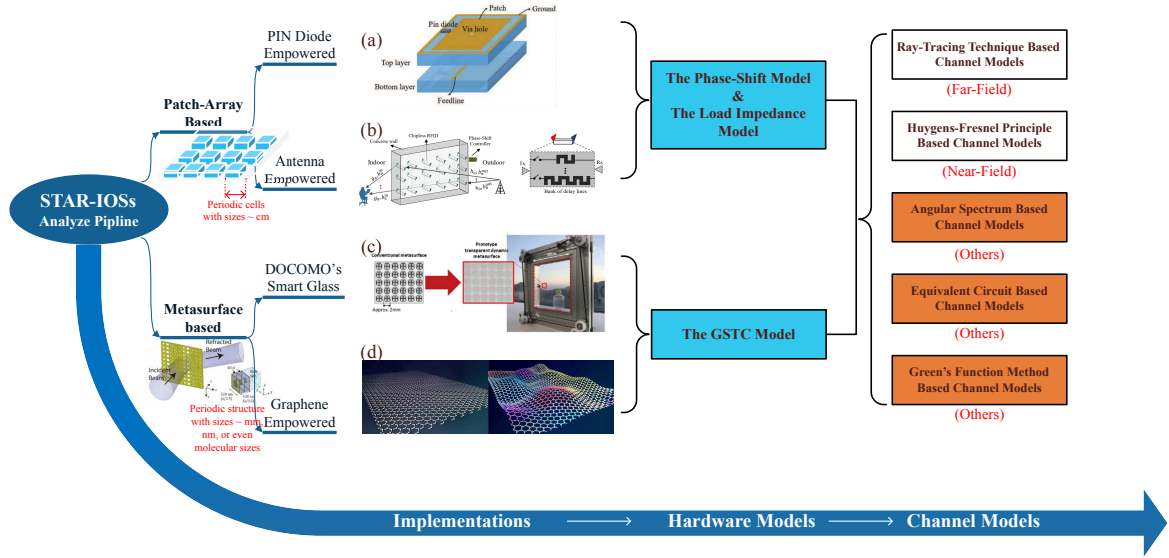}
        \caption{Framework for analyzing STAR-IOSs. (a) PIN diode empowered STAR-IOS~\cite{IOS}, (b) antenna empowered STAR-IOSs~\cite{oti}, (c) DOCOMO's smart glass~\cite{DOCOMO}, (d) graphene empowered STAR-IOSs.}
        \label{fig.big}
    \end{center}
\end{figure*}

While STAR-IOSs attain various benefits, such as
improving the coverage of the SRE, this is achieved at the cost of 
additional hardware requirements and 
the feasibility of implementing STAR-IOSs is not well documented. More specifically, the following fundamental questions have to be answered:
\begin{itemize}
\item \textit{Q1}: How can STAR-IOSs be implemented?
\item \textit{Q2}: Can STAR-IOSs achieve independent control of both the reflected and transmitted signals?
\item \textit{Q3}: How can the phase shift or amplitude change that a STAR-IOS imposes on the reflected and transmitted wireless signals be characterized?
\item \textit{Q4}: How can the received signal power and performance gain of receivers at both sides of the STAR-IOS be calculated?
\end{itemize}

To answer these questions comprehensively, we propose a general framework for the analysis of STAR-IOSs. The main contributions are as follows:
\begin{itemize}
\item We summarize four promising techniques that may be used for implementing STAR-IOSs, which also allow the independent control of both the reflected and transmitted signals.
\item We present three hardware models for STAR-IOSs. These models characterize the tuning capability of STAR-IOSs with different levels of accuracy.

\item We categorize the associated channel models into five types, one for calculating the far-field channel gain, one for the near-field, as well as three other physics-based channel models. The pros and cons of these channel modeling methods are also discussed.
\end{itemize}

As illustrated in Fig~\ref{fig.big}, our study of the STAR-IOS relies on three steps. 
One tangible benefit of our approach is the resultant clear distinction between the imperfections of the hardware model and those of the channel model, as discussed further below.

\section{Hardware Implementations for STAR-IOSs}\label{sec:imp}
How the \textit{STAR} concept can be implemented based on practical hardware designs is one of the most pivotal questions in research of STAR-IOSs. To address this issue, in this section, we survey and categorize the possible implementation options of STAR-IOSs to achieve independent control of the reflected and transmitted signals.

There are various tunable surface designs which are potential candidates for realizing STAR-IOSs. In \cite{ahead}, the authors pointed out an intuitive difference between natural and artificial materials (RISs in general), namely that natural materials exhibit an uniform EM properties along their tangential directions, while artificial materials exhibit either a periodic or quasi-periodic nature.
In terms of the periodic structure, we can loosely classify the hardware implementations of STAR-IOSs into two categories, namely the patch-array based implementations and the metasurface based implementations.
%
As illustrated in Fig.~\ref{fig.big}, the patch-array based implementations consist of periodic cells having sizes on the order of a few centimetres. Because of their relatively large sizes, each cell (patch) can be made tunable by incorporating either PIN diodes or delay lines. By contrast, the matasurface based implementations have periodic cells on the order of a few millimetres, possibly micrometres, or even molecular sizes. Hence they require more sophisticated controls of their EM properties, such as the conductivity and permittivity. Below, we provide a brief overview of a pair of patch-array based implementations and two metasurface based implementations. All these hardware implementations have had successful prototypes built or rely on strong theoretical evidence in support of their feasibility.

\begin{table*}[!t]
\caption{Comparing different implementations of STAR-IOSs.}\label{table_last}
\centering
\begin{tabular}{|c|c|c|c|c|}
\hline
Implementations                                                              & Operating frequency                                                                                        & STAR-IOS prototypes  & Tuning mechanism  & Independent reflection/transmission control         \\ \hline
\multirow{2}{*}{\begin{tabular}[c]{@{}c@{}}Patch-array\\ based\end{tabular}} & \multirow{2}{*}{\begin{tabular}[c]{@{}c@{}}Low to high frequency\\ (10KHz up to 1GHz)\end{tabular}}        & PIN diode empowered  & Bias voltages on PIN diodes & Difficult to achieve \\ \cline{3-5} 
                                                                             &                                                                                                            & Antenna empowered    & Lengths of delay lines  & Can be achieved    \\ \hline
\multirow{2}{*}{\begin{tabular}[c]{@{}c@{}}Metasurface\\ based\end{tabular}} & \multirow{2}{*}{\begin{tabular}[c]{@{}c@{}}Super high frequency\\ to visible light frequency\end{tabular}} & DOCOMO's smart glass & Distance between substrates & Theoretically achievable \\ \cline{3-5} 
                                                                             &                                                                                                            & Graphene empowered   & Conductivity of graphene  & Can be achieved  \\ \hline
\end{tabular}
\end{table*}

\subsection{Patch-array Based STAR-IOSs}
\subsubsection{PIN Diode Empowered Implementations}
For patch-array based implementations, both the ph	ase and amplitude response
can be tuned by applying different bias voltages to the positive intrinsic negative (PIN) diodes. In \cite{IOS}, the authors presented a STAR-IOS prototype relying on the PIN diode empowered implementation. This implementation is the most popular design for both RISs and STAR-IOSs since PIN diodes are of low-cost and are voltage-controlled. The drawback of this implementation is that since PIN diodes only have two states, namely, ``ON'' or ``OFF'', this implementation can only support a finite-cardinality reflection and transmission coefficient set. Moreover, for a given state of all the PIN diodes, the reflection and transmission coefficients are coupled. As a result, the PIN diode empowered implementation struggles to mimic independent control of both reflection and transmission unless a sufficiently high number of PIN diodes are used for each patch element.

\subsubsection{Antenna Empowered Implementations}
The concept of phased-array antennas may be readily extended to STAR-IOSs
with some minor modifications. As illustrated in Fig.~\ref{fig.big}(b), according to \cite{oti}, each antenna empowered STAR-IOS cell actually consists of two antenna elements, which are connected by a tunable delay line (waveguide). 
The antenna elements facing the incident wave operate similarly to the reflecting-only RIS elements but a certain fraction of the incident energy is transferred along the delay lines and it is re-radiated into the \textit{transmission space}. The phase of the transmission coefficient of each cell is determined by the length of the delay line. Thus, the phase shifts of both the reflected and the transmitted signals can be independently adjusted.
However, the drawback of this implementation is that the delay line may impose a considerable energy losts. If the desired phase shift of the transmitted signal is high, the amplitude of the transmission coefficient will be reduced. Thus, the amplitude and phase of the transmission coefficient are correlated in this implementation.

\subsection{Metasurface Based STAR-IOSs}

\subsubsection{DOCOMO's Smart Glass}

A popular prototype of the metasurface based STAR-IOSs is the \textit{transparent dynamic metasurface} designed by researchers at NTT
DOCOMO, Japan (Fig.~\ref{fig.big}(c)). According to \cite{DOCOMO}, the metasurface supports the manipulation of 28 GHz (5G) radio signals. It allows dynamic control of both of the signal's reflection and transmission while maintaining transparency of the window. 
By adjusting the width of the dielectric material and its distance between substrates, the smart glass can be switched into different modes, such as full penetration (transmission), full reflection, and partial reflection.
DOCOMO revealed that they are working on more sophisticated tuning techniques and will use them in future trials. 

The main advantage of DOCOMO's smart glass is that owing 
to its transparency in the visible light frequency range, it can be aesthetically integrated into buildings. However, its drawback is that it does not have the ability to dynamically reconfigure itself as the PIN diode empowered implementations. Moreover, adjusting the distance between substrates may affect the reflection coefficient of the entire surface instead of only reconfiguring a particular element.

\subsubsection{Graphene Empowered Implementations}
It has been widely exploited that
a single graphene layer has extraordinary properties, including a beneficial EM wave response that may also be used for building STAR-IOSs. More importantly, to realize futuristic envisions such as smart surfaces assisted visible light wireless transmission and wearable skin-like smart surfaces, we might have to relay on graphene empowered technologies. 
Indeed, there are already experimental graphene-based RF devices~\cite{gra_1}. To achieve reconfigurability, a single layer of graphene has tunable reflection and transmission coefficients by adjusting its conductivity. Moreover, a periodic stack of graphene layers is capable of acting as a tunable spectrally-selective mirror. We may summarize that for graphene empowered STAR-IOSs, even the separation of a combined signal might become feasible based on the different carrier frequencies or polarizations. 
In a nut shell, graphene empowered implementations might open extraordinary possibilities for the design of smart radio environments once their fabrication process becomes more economical.

\begin{figure*}[!t]
\centering
\subfigure[The phase-shift model]{\label{2a}
\includegraphics[scale=0.5]{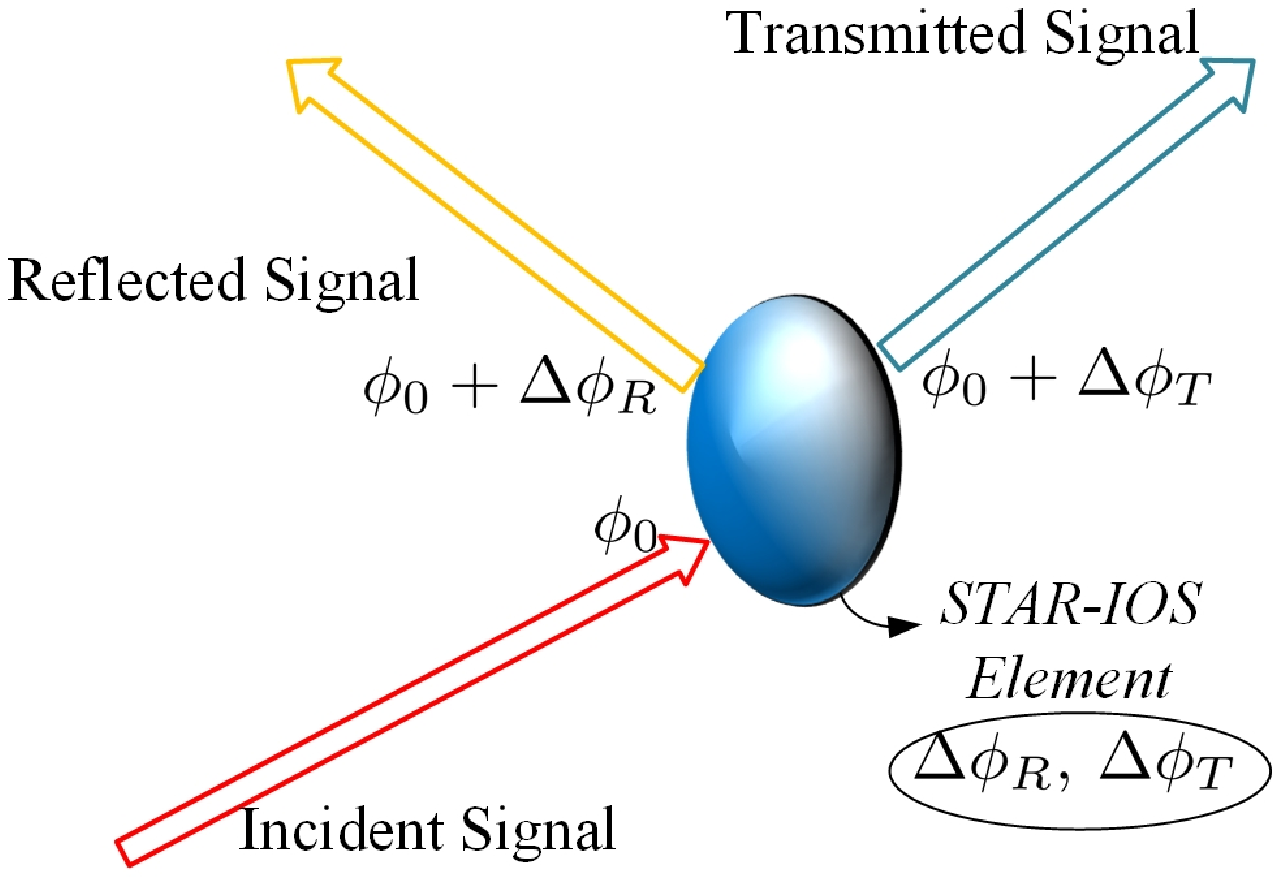}}
\subfigure[The load impedance model]{\label{2b}
\ \ \ \ \ \ \ \ \ \ \ \ \ \ 
\includegraphics[scale=0.5]{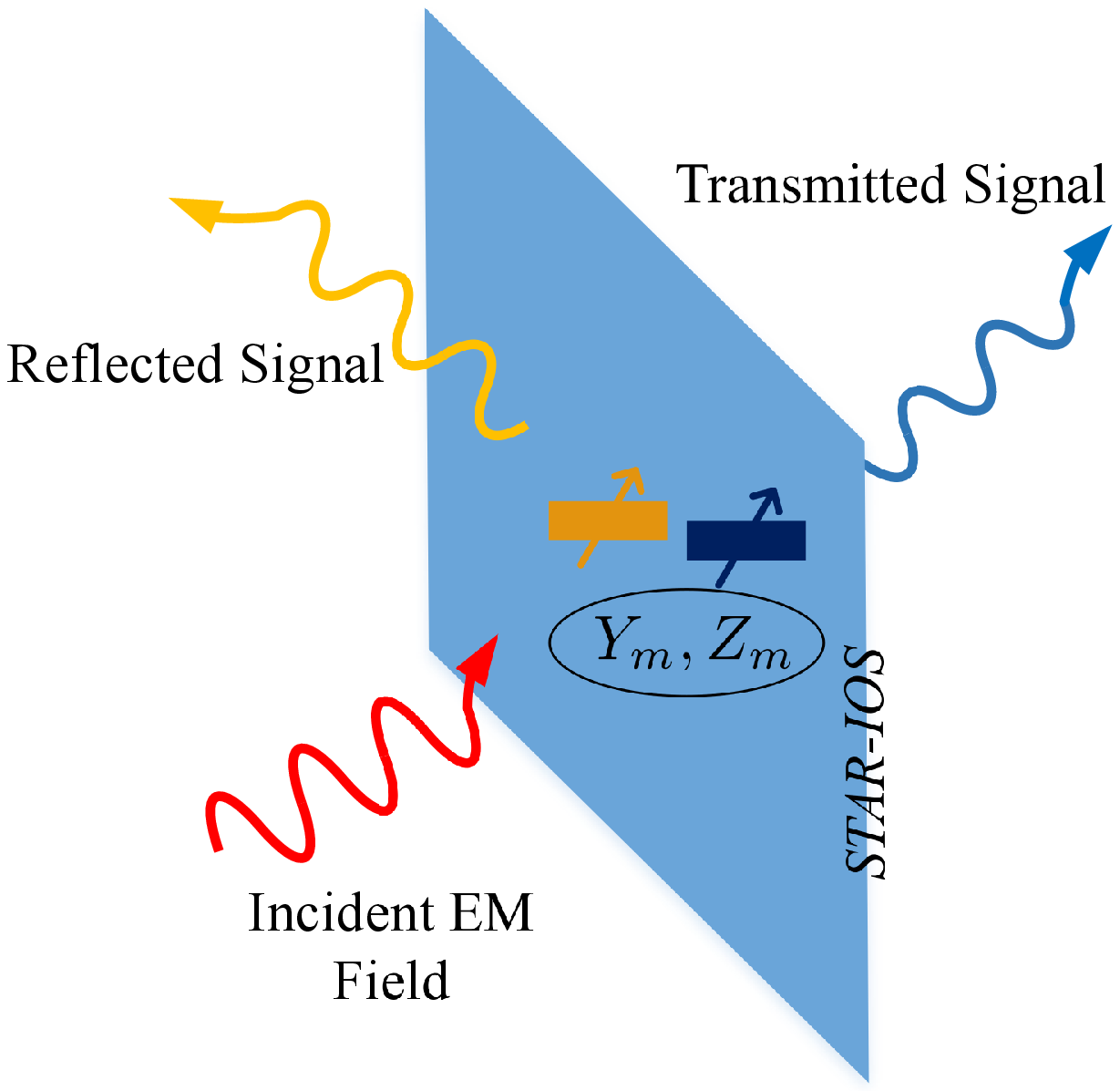}}
\caption{Conceptual comparison between different models for the patch-array based STAR-IOS implementations.}\label{fig_new}
\end{figure*}

\subsection{STAR-IOS Implementations and Operating Frequencies}
Naturally, almost all designs can only operate as desired within a certain frequency range. This is because in order for the STAR-IOS to apply the desired phase shifts and wave-front transformations both to the transmitted and reflected signals, the length of periodicity in STAR-IOSs has to \textit{match} the wavelength of the wireless signal. For the patch-array based implementations, the periodicity is usually chosen to be between $0.5\lambda$ to $0.7\lambda$, where $\lambda$ is the wavelength of the wireless signal~\cite{array_handbook}. According to this relationship, the patch-array based STAR-IOSs are suitable for assisting wireless communication up to $1$GHz carrier frequency. For wireless signal having higher frequency and for visible light communication, metasurface based STAR-IOSs are required.

In conclusion, all the above-mentioned STAR-IOS implementations achieve independent control of both the reflected and transmitted signals when an adequate number of control units is applied to each element. However, there is a trade-off between the phase/amplitude control accuracy and design complexity. 
This is reflected in the increase of both the energy consumed per adjustment and the time delay of each adjustment. In Table~\ref{table_last}, we summarize the operating frequency and tuning mechanism of each STAR-IOS implementation discussed.


\section{Hardware Models for STAR-IOSs}\label{sec:hard}

As discussed in Section.~\ref{sec:imp}, different STAR-IOS implementations are rather different in terms of their tuning mechanisms. However, we require a unified technique of modeling the effect of these surfaces on the wireless signal. Explicitly, we have to find an accurate hardware model for characterizing the EM wave response of the STAR-IOSs.
From a tangible physical perspective, modeling a smart surface is equivalent to the problem of studying the boundary conditions of the EM field at the surface. However, the interaction of an arbitrary field with the STAR-IOS is an intrinsically complex, all existing hardware models use different approximations and assumptions. Our next objective is to demonstrate and compare these assumptions, as well as to reveal the physical abstractions behind each model.

\subsection{The Phase-Shift Model}

The phase-shift model characterizes a smart surface using a collection of phase shift values, or applying a specific phase shift as a function of the cell's position on the surface. This function is also often referred to as the phase profile, phase discontinuity, or phase-shift matrix, depending on the context. As illustrated in Fig.~\ref{2a}, the physical abstractions laying the foundation of the phase-shift model are as follows: The STAR-IOS can be regarded as a periodic array of either metallic or dielectric particles. Regardless of the specific geometric and electromagnetic properties of these particles, the reflected or transmitted field radiated from the STAR-IOS can be
characterized by the superposition of waves
radiated from different particles, each having a phase delays induced by the corresponding particle. As a result, the only hardware features of this model are the positions of the particles, i.e., the STAR-IOS elements and their corresponding phase shifts associated with the reflection and transmission, i.e., $\Delta\phi_R$ and $\Delta\phi_T$ in Fig.~\ref{2a}. 
The phase-shift model is widely adopted and convenient to use. However, it is an over-simplified representation of the actual physical process. As a result, it cannot accurately characterize either the energy flow at the surface or the non-local power transfer effects~\cite{ahead}.

\subsection{The Load Impedance Model}

In the load impedance model, each element is modeled as a lumped circuit having surface-averaged electric and magnetic impedances of $Y_m$ and $Z_m$. As illustrated in Fig.~\ref{2b}, the physical processes of wave reflection and transmission may be portrayed as follows: each element of the STAR-IOS is excited by the incident wave. After being excited, both electrical and magnetic currents are induced, whose intensity depends on the effective voltage of the incident wave and the equivalent load impedance of the circuit element. Finally, the currents induced generate an EM field, which is radiated towards both sides of the STAR-IOS. As a result, the hardware features of the model are the position and load impedances of each passive element. The problem of determining the field radiated by the currents flowing through each element is left to deal with by the channel model.
The load impedance model can be reduced to the phase-shift model by incorporating some further idealized simplifying assumptions because both 
the reflection and transmission coefficients of the surface can be formulated as a function of the surface impedances. Specifically, the reflection and transmission coefficients of the $m$th element are defined as the ratio between electric fields, which may be represented by complex numbers~\cite{xu_star}.
The argument of the reflection and transmission coefficients for each element correspond to the phase delay values in phase-shift hardware model. In light of this, the phase-shift model can be regarded as a simplified version of the load impedance model.

\subsection{The Generalized Sheet Transition Conditions Model}

The generalized sheet transition conditions (GSTC) model~\cite{ahead} is the most general one of the three hardware models discussed because it is based on a continuous distribution of the electric and magnetic polarization densities of the surface, instead of relying on a finite number of impedance values.
The GSTC model uses the electric and magnetic susceptibilities as a function of position on the surface for characterizing the smart surface.
According to Maxwell's equations, the conventional boundary conditions at the surface describe the discontinuity of the EM field in terms of the surface electric and magnetic polarization densities. In the GSTC model, these surface electric and magnetic polarization densities are induced by the incident field and depend on the polarizability densities of the material. As a result, the reflected and transmitted field can be formulated using only material-dependent electric and magnetic polarizability dyadics~\cite{ahead}. Moreover, the electric and magnetic polarizabilities of the scatterers are microscopic properties of the material, thus the GSTC model is capable of describing the metasurface based STAR-IOS implementations relying on small periodic structures. At the same time, the GSTC model can also be used for modeling patch-array based implementations by taking the surface-average of the electric and magnetic polarizabilities within each element.

\section{Channel Models Based On Electromagnetic Arguments}\label{sec:channel}
Once the hardware model has been selected,
the final step to reconcile physical implementations with communication theory is to adopt a channel model for determining the received signal power.
There are various physics-based approaches that can be used to develop channel models for STAR-IOSs~\cite{green,angular,end-to-end},
hence below we critically appraise
five promising approaches\footnote{We would like to point out that there are other existing research works on the topic of RIS channel modeling. Those works studied the RIS channel in specific application scenarios based on the conventional ray tracing method. Since we focus on novel physic-based channel models for STAR-IOSs in this paper, the contribution of those works are not discussed here.}.
\begin{figure}[t!]
    \begin{center}
        \includegraphics[scale=0.6]{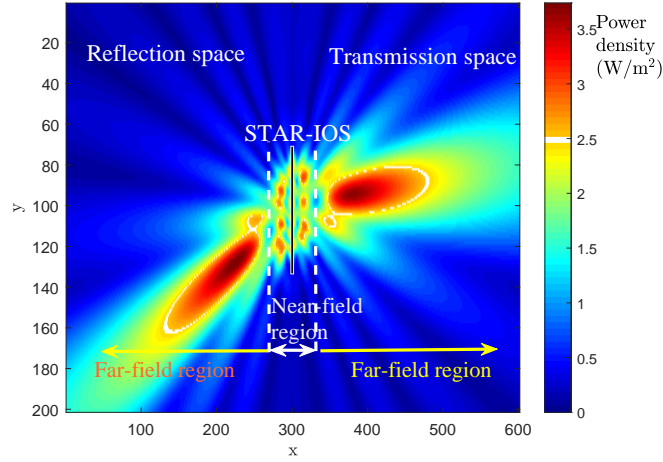}
        \caption{Simulated radiation pattern and field regions for a $10\times 10$-element patch-array based STAR-IOS.}
        \label{rad}
    \end{center}
\end{figure}
\subsection{Ray-Tracing Based Channel Models}
One of the most widely-accepted far-field channel models is the ray-tracing based one~\cite{xu_star}.
The ray-tracing technique has long been adopted as an efficient way to simplify the calculation of wave propagation and obtain the channel gains of receivers located within the far-field region\footnote{For more technical elaboration on the far-field and near-field channel models and the boundary between these two field regions, please refer to \cite{xu_star}.}. In addition, the conventional ray-tracing technique is only compatible with the phase-shift based STAR-IOS hardware model or the load impedance model. This is because the ray-tracing technique assumes a finite number of scatters and studies their sum, hence it cannot deal with continuous phase profiles.

In the context of analyzing the STAR-IOSs, the ray-tracing technique relies on the following assumptions: 
\begin{enumerate}
\item Each element of the STAR-IOSs is treated as a distinct scatter having a known location and dielectric properties.
\item The wave impinging on the STAR-IOS is regarded as a collection of rays, each falling on a single element. Thus, the received bundle of rays is constituted by a discrete 2-D array instead of a EM field in 3-D space.
\item The interactions between each ray and each element, including the reflection and transmission, are studied using geometrical optics instead of wave optics. 
\end{enumerate}

\begin{table*}[!t]
\centering
\caption{Comparing different channel models for STAR-IOSs.}\label{table2}
\begin{tabular}{|l|r|r|r|}
\hline
Channel Models & Advantages & Disadvantages & Challenges \\ \hline

\multirow{2}{*}{Ray-tracing based models}& Simplistic in form and   &  only apply under certain           &         Describing the channel in         \\ 
& easy to calculate channel gain & conditions in far-field regions &the near-field regions \\ \hline

\multirow{2}{*}{Huygens-Fresnel principle based models} & Fundamental and applies & Apply only to free-space       & Choosing proper wave-front     \\ 
&to near-field regions & scenarios or LoS-dominate links & to characterize the STAR-IOS\\ \hline
\multirow{2}{*}{Angular spectrum based models} & Convenient for designing           & Apply only to free-space     & Deciding the boundaries between     \\
&desired aperture distributions & scenarios or LoS-dominate links &  different field regions\\ \hline

\multirow{2}{*}{Equivalent circuit based models}& Simplistic in form and   &  only apply for linear          &         Describing the channel for system        \\ 
& easy to calculate channel gain & system in free-space  & with non-linear filters \\ \hline

\multirow{2}{*}{Green's function method based models} & Fundamental and apply & Complex and requires detailed  & Choosing proper boundary conditions     \\ 
&to general cases&system specifications& to characterize the environment \\ \hline

\end{tabular}

\end{table*}

\subsection{Huygens-Fresnel Based Channel Models}
\begin{figure}[ht!]
    \begin{center}
        \includegraphics[scale=0.4]{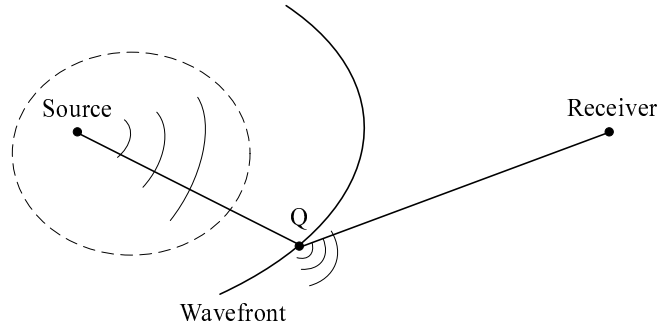}
        \caption{Illustration of the Fresnel-Kirchhoff diffraction formula.}
        \label{fres}
    \end{center}
\end{figure}

The Huygens-Fresnel principle is method to solve the problems of wave propagation. It applies to both the far-field region and the near-field region.  It states that every point on a wavefront is itself the source of spherical wavelets, and the sum of these spherical wavelets forms the wavefront. According to this principle, A. Fresnel and Kirchhoff. G arrived at the analytical result which is known as the Fresnel-Kirchhoff diffraction formula. This formula is the theoretical foundation to calculate the channel gain in Hyugen-Fresnel principle based channel models. As illustrated in Fig.~\ref{fres}, the chosen wavefront is represented with a solid curve. In real applications, the chosen wavefront should be a 2-D plane, preferably located in the same plane as the RIS. In spirit of the Huygens-Fresnel principle, the EM field at the receiver can be calculated as the sum of contributions of each point on the chosen wavefront. According to the Fresnel-Kirchhoff diffraction formula, the contribution to the received field at each wave point is proportional to its area on the wavefront, the amplitude of the corresponding wavelet, the leaning factor at each point on the wavefront, and the inverse of the distance between each point on the wavefront and the receiver. In light of this, one can immediately notice that the received field can be completely characterized by the field at the wavefront. In other words, the information related to the source (illustrated as the region inside the dashed circle in Fig.~\ref{fres}) can be equivalently described using the wavefront. Thus, an important difference between the Green's function method based channel models and Huygens-Fresnel principle based channel models can be drawn: the Huygens-Fresnel principle based channel models calculate the end-to-end channel gain base on the field distribution on a chosen wavefront while the Green's function method based channel models calculate the channel gain base on sources. This is to say one model only studies the propagation of the wave, the other studies the generation of the wave. Base on this observation, it is easy to notice that the Huygens-Fresnel principle based channel model is the best choice for analyzing the wavefront transformation function of the RIS. For example, if the incident field is a plane wave, then its wavefront evaluated at any plane should be of a uniform amplitude and with a linear phase distribution. Suppose we adopt the simplest phase-shift RIS hardware model, a wavefront transition from a plane wave to another plane wave can be achieved by implementing a linear phase gradient on the RIS. By using the Huygens-Fresnel principle, it can be easily confirmed that the reflected field is a plane wave radiated at a direction related to the RIS phase gradient and the angle of incident plane wave.

Similar with the Green's function method based channel models, in general cases, the Huygens-Fresnel principle based channel models present the end-to-end channel gain in integral form. The integral typically runs over the surface area of the RIS because the area outside the RIS on the chosen wavefront is assumed to have no contribution to the received signal. This assumption is only true for the reflective RIS in free-space. For the case of refractive RIS where the transmitter and the receivers are located at different sides of the RIS, the integral must be extended to a complete closed wavefront. Despite of the many advantages of the Huygens-Fresnel principle based channel models, they apply the best only for the free-space systems. For wireless networks involving both RIS and other uncontrollable scatters in the environment, we need to relay on the Green's function method base channel models.

\subsection{Green's Function Method Based Channel Models}
\begin{figure}[ht!]
    \begin{center}
        \includegraphics[scale=0.2]{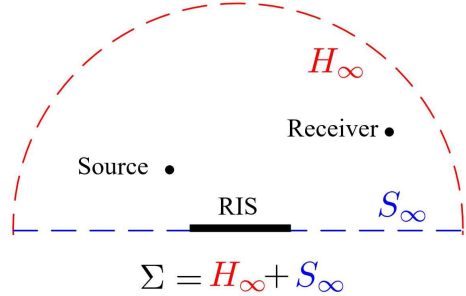}
        \caption{Closed boundary $\Sigma$ for the boundary condition in Helmholtz problems, reproduced from \cite{ahead}, Fig. 2.}
        \label{boundary_con}
    \end{center}
\end{figure}

The Green's function method is a mathematical approach to solve inhomogeneous linear differential equations. It is relevant to the discussion of RIS-assisted channel because the electromagnetic field in front of the RIS, according to Maxwell's equations, satisfies the inhomogeneous Helmholtz equation. The Helmholtz equation describes the fundamental property of a sinusoidal time-dependent EM wave. The Helmholtz equation is essentially equivalent to the wave function which link the spatial derivative, the time derivative of the field with the source. In addition, to fully determine the field, a boundary condition is need. As a illustrated in Fig.~\ref{boundary_con}, the closed surface $\Sigma$ is a proper choice of the boundary since it encloses the space where the receiver locates. This closed surface is formed by an infinite plane at the RIS and a hemisphere with an radius tends to infinity. With this chosen closed boundary, the boundary condition can be expressed as the complex value of the EM field at the surface, which is known as the Dirichlet boundary condition, or as the derivative of the EM field along the normal direction of the surface, which is the Neumann boundary condition.
The Helmholtz equation together with the boundary condition determine the radiating field at any point inside boundary $\Sigma$. Thus, the received radio power can be derived using the power density of the radiating field and the specifics of the receiving antenna's gain.

The critical challenge of Green's function method is how to chose the proper boundary condition. Since all the information carried by the RIS is represented in boundary conditions. 
In light of this, the Green's function based channel models are more compatible with the GSTC model since it gives more detailed boundary conditions.
At the time of writing, proposed Green's function based channel models are based on free-space scenario. For application scenario where obstacles and scatters are involved, different shapes of boundary surface need to be used to characterize their effects.
The advantages of the Green's function method based channel models are as follows: First, it is a general and physic-compliant model which gives physically meaningful predictions of the end-to-end channel gain. Secondly, it capture the EM wave response of the RIS as well as the environment in a detailed way. As a result, it can reflect physical phenomena that can not be characterized by naive models, such as the polarizations of EM waves, the mutual coupling effect and non-local behaviour of the RIS elements, and the radiation pattern in the near-field region. At the same time, the cost is that the resulting channel gain can only be presented by integrals in Green's function method based channel models. However, by further adopting proper approximations in different scenarios, the integrals can be reduced to more tractable forms.

In conclusion, the Green's function method based models are the best choices of channel models when the application scenario require detailed inspection of the RIS hardware. In addition, this type of channel models are the most compatible in system models where the transmitters are characterized by equivalent source of electric and magnetic currents or dipoles. For system models in which only the wave forms radiated from the transmitters are of interests, the next few models are more appropriate to be adopted in these situations.

\subsection{Angular Spectrum Based Channel Models}\label{angular}
They also rely on the Huygens-Fresnel principle. However, instead of calculating the channel gain using the Fresnel-Kirchhoff diffraction formula, the model exploits the fact that the EM fields on both sides of the STAR-IOS having any arbitrary distributions can be regarded as a collection of plane waves travelling in different directions. In light of this, 
provided that 
the spectrum of this collection of plane waves can be determined, the channel gain can also be derived at any point. In angular spectrum based channel models, the wavefront is chosen as the plane in which the STAR-IOS lies. In the field of antenna design, the 2-D spatial distribution of the EM field at this chosen wavefront is referred to as the \textit{aperture distribution}~\cite{array_handbook}, characterizing the complex-valued amplitude as a function of position on the smart surface. An essential statement that empowered the angular spectrum based channel models is that the plane-wave angular spectrum of an arbitrary wave form can be given by the Fourier transform of the aperture distribution~\cite{angular}. In general, an arbitrary aperture distribution can be expressed as a Fourier expansion of a series of plane waves having different wave numbers. Correspondingly, the radiation obeying this aperture distribution can be expressed as a spectrum of plane waves. More detailed inspection shows that not all the wave numbers in the Fourier expansion of the aperture distribution correspond to propagating waves. In fact, aperture distributions having large wave number values give rise to \textit{evanescent waves}. These waves does not propagate well, because they decay exponentially with distance, hence they do not usefully contribute to the received power beyond a few wavelengths. As shown in Fig.~\ref{rad}, the field region affected by these evanescent waves is termed as the reactive near-field region.

\subsection{Equivalent Circuit Based Channel Models}
Instead of studying the propagation or generation of the EM wave, the equivalent circuit based channel models characterize the channel between the STAR-IOS and the receivers using a linear transformation~\cite{end-to-end}. Explicitly, in this model, each element and receiver is represented by specific \textit{ports} of the circuit. The overall circuit consists of a collection of load impedances. The total number of ports is equal to the number of STAR-IOS elements plus the number of receiving antennas. Each port has different voltages and hence carries different current. The load impedance matrix relates these voltages and currents to each other by a linear transformation. Thus, at the receiving ports, the time-averaged power can be calculated as the product of the current and voltage. However, in general cases, the EM field response of the elements is not linear. As a result, we believe that the validity of the model in general scenarios has to be further justified.

In conclusion, in Table~\ref{table2}, all five channel models are summarised in terms of their advantages, disadvantages, and challenges, as the main takeaway message of the paper.

\section{Conclusions}
In this paper, the implementation and modeling of STAR-IOSs was presented. Specifically, we pointed out that STAR-IOSs can can be implemented relying on both patch-array based and on metasurface based technologies. Successful prototypes and hardware models of the STAR-IOS were also presented to illustrate how independent control of the reflected and transmitted signals can be achieved.
Then, channel models with different levels of accuracy were summarized and compared. Since the STAR-IOS concept is capable of supporting multiple users, future directions of studying STAR-IOSs include the research of their practical operating protocols, designing multiple access schemes, and finding further application scenarios.


\begin{thebibliography}{9}

\bibitem{pan}
C. Pan, \textit{et al.}, ``Reconfigurable Intelligent Surfaces for 6G Systems: Principles, Applications, and Research Directions,'' \textit{IEEE Commun. Mag.}, vol. 59, no. 6, pp. 14-20, June 2021.

\bibitem{ahead} 
M. Di Renzo, \textit{et al.}, ``Smart Radio Environments Empowered by Reconfigurable Intelligent Surfaces: How It Works, State of Research, and The Road Ahead,'' \textit{IEEE J. Sel. Areas Commun.}, vol. 38, no. 11, pp. 2450-2525, Nov. 2020.

\bibitem{liu_survey} 
Y. Liu, \textit{et al.}, ``Reconfigurable  intelligent  surfaces:  Principles  and opportunities,'' \textit{IEEE  Commun.  Surv.  Tutor.},  Early  Access,  2021,  doi:10.1109/COMST.2021.3077737.

\bibitem{STAR_mag}
Y. Liu, \textit{et al.}, ``STAR: Simultaneous Transmission And Reflection for 360$^\circ$ Coverage by Intelligent Surfaces,'' arXiv preprint arXiv:2103.09104, 2021.

\bibitem{IOS}
H. Zhang, \textit{et al.}, ``Intelligent Omni-Surfaces for Full-Dimensional Wireless Communications: Principle, Technology, and Implementation,'' {arXiv preprint arXiv:2104.12313}, 2021.

\bibitem{xu_star}
J. Xu, Y. Liu, X. Mu and O. A. Dobre., ``STAR-RISs: Simultaneous Transmitting and Reflecting Reconfigurable Intelligent Surfaces,'' \textit{IEEE Communications Letters,} to appear.

\bibitem{Xidong}
X. Mu, \textit{et al.}, ``Simultaneously transmitting and reflecting (STAR) RIS aided wireless communications.'' arXiv preprint arXiv:2104.01421, 2021.

\bibitem{shuhang}
S. Zhang, \textit{et al.}, ``Intelligent Omni-Surface: Ubiquitous Wireless Transmission by Reflective-Refractive Metasurface,'' \textit{IEEE Trans. Wireless Commun.,} to appear.




\bibitem{oti}
M. Nemati, \textit{et al.}, ``Modeling RIS Empowered Outdoor-to-Indoor Communication in mmWave Cellular Networks,'' {arXiv preprint arXiv:2101.00736}, 2021.

\bibitem{DOCOMO}
NTT DOCOMO, ``DOCOMO Conducts World's First Successful Trial of Transparent Dynamic Metasurface,'' Jan. 2020. [Online]. Available:
\url{nttdocomo.co.jp/english/info/media\_center/pr/2020/0117\_00.html}.



\bibitem{gra_1}
W. Zhu, \textit{et al.}, ``Graphene Radio Frequency Devices on Flexible Substrate,'' \textit{Appl. Phys. Lett.}, 2013, 102(23): 233102.



\bibitem{array_handbook} 
R. J. Mailloux, \textit{Phased Array Antenna Handbook}. Artech house, 2017.

\bibitem{end-to-end}
G. Gradoni and M. Di Renzo, ``End-to-End Mutual Coupling Aware Communication Model for Reconfigurable Intelligent Surfaces: An Electromagnetic-Compliant Approach Based on Mutual Impedances,'' \textit{IEEE Wireless Communications Letters}, vol. 10, no. 5, pp. 938-942, May 2021.
\bibitem{green}
F. H. Danufane, M. Di Renzo, J. De Rosny and S. Tretyakov, ``On the Path-Loss of Reconfigurable Intelligent Surfaces: An Approach Based on Green’s Theorem Applied to Vector Fields,'' \textit{IEEE Transactions on Communications}, to appear.

\bibitem{angular}
H. G. Booker, and P. C. Clemmow, ``The concept of an angular spectrum of plane waves, and its relation to that of polar diagram and aperture distribution,'' \textit{Proceedings of the IEE-Part III: Radio and Communication Engineering}, 97.45 (1950): 11-17.

\end{thebibliography}

\bibliographystyle{IEEEtran}

\end{document}